\documentclass[aps,pre,preprint,superscriptaddress]{revtex4-1}
\usepackage{graphicx}
\usepackage{amsmath}
\usepackage{amssymb}
\usepackage{psfrag}
\usepackage{natbib}

\newcommand{\pd}[2]{\displaystyle\frac{\partial #1}{\partial #2}}
\newcommand{\ppd}[2]{\displaystyle\frac{\partial^2 #1}{\partial #2^2}}

\newcommand {\e} {\varepsilon}

\def \w {\omega}
\def \W {\Omega}
\def \e {\varepsilon}
\def \vp {\varphi}

\def \ii {\text{i}}

\newcommand{\be}{\begin{equation}}
\newcommand{\ee}{\end{equation}}

\newcommand{\bea}{\begin{eqnarray}}
\newcommand{\eea}{\end{eqnarray}}

\begin{document}
	\title{Synchronization of oscillators in a Kuramoto-type model with generic coupling}

\author{Vladimir Vlasov}  \email{mr.voov@gmail.com}
\affiliation{Department of Physics and Astronomy, Potsdam University, 14476
Potsdam, Germany}
\author{Elbert E. N. Macau}
\affiliation{National Institute for Space Research - INPE, 12227-010 
Sao Jose dos Campos, SP, Brazil}
\author{ Arkady Pikovsky}
\affiliation{Department of Physics and Astronomy, Potsdam University, 14476
Potsdam, Germany}   
\affiliation{Department of Control Theory, Nizhni Novgorod State 
University, Gagarin Av. 23,
606950, Nizhni Novgorod, Russia}
 \begin{abstract}
We study synchronization properties of coupled oscillators on networks
that allow description in terms of global mean field coupling. These models
generalize the standard Kuramoto-Sakaguchi model, allowing for different 
contributions of oscillators to the mean field and to different forces from
the mean field on oscillators. We present the explicit solutions of
self-consistency equations for the amplitude and frequency
of the mean field in a parametric form, valid for noise-free and noise-driven
oscillators. As an example we consider spatially spreaded oscillators, for which
the coupling properties are determined by finite velocity of signal propagation.
 \end{abstract}
	\date{\today}

	\maketitle

\textbf{
Synchronization of large ensembles of oscillators is an ubiquitous phenomenon in
physics, engineering, and life sciences. The most simple setup pioneered by Winfree and Kuramoto
is that of global coupling, where all the oscillators equally contribute to a mean
field which acts equally on all oscillators. 
In this study we consider a generalized Kuramoto-type model of mean field coupled oscillators 
with different parameters for all elements. In our setup there is still a unique
mean field, but oscillators differently 
contribute to it with their own phase shifts and coupling factors, and also 
the mean field acts on each oscillator with different phase shifts and coupling coefficients. 
Additionally, the noise term is included in the consideration. 
Such a situation appears, e.g., if the oscillators are spatially arranged and 
the phase shift and the attenuation due to propagation
of their signals cannot be neglected.
A regime, where the mean field rotates uniformly, 
is the most important one. For this case the solution of the self-consistency equation for 
an arbitrary distribution of frequencies and coupling parameters is found analytically 
in the parametric form, both for noise-free and noisy oscillators. First, we consider 
independent distributions for the coupling parameters when self-consistency equations can be 
greatly simplified. Secondly, we consider an example of a particular
geometric organization of 
oscillators with one receiver that collects signals from oscillators, 
and with one emitter 
that sends the driving field on them.  By using our approach, synchronization 
properties can be found for 
different geometric structures and/or for different parameter distributions. }
	
\section {Introduction}
Kuramoto model of globally coupled phase oscillators lies at the basis 
of the theory of  synchronization of oscillator
populations~\cite{Kuramoto-84,Acebron-etal-05}.
The model can be formulated as the maximally homogeneous
mean field interaction:
all oscillators equally contribute to the complex mean field, and this field 
equally acts on each oscillator (when this action also includes a phase shift,
common for all oscillators,
one speaks
of the Kuramoto-Sakaguchi model~\cite{Sakaguchi-Kuramoto-86}). The only
complexity in this setup stems
from the distribution of the natural frequencies of the oscillators, and from
a possibly nontrivial 
form of the coupling function (which can be e.g. a nonlinear function of the mean 
field~\cite{Rosenblum-Pikovsky-07,Pikovsky-Rosenblum-09}). 

If one considers coupled oscillators on networks, quite a large variety of
setups is possible
where different oscillators are subject to different inputs, so that
mean fields are not involved in the interaction and thus the coupling
cannot be described
as a global one. In this paper we consider a situation where the oscillators are
structured as a specific network that allows one to describe the coupling as a
global one.
We assume that there is some complex ``global field'' which involves
contributions from
individual oscillators, and which acts on all of them. However, 
contrary to the usual
Kuramoto-Sakaguchi setup,
we assume the contributions to the global field to be generally different,
depending on 
individual oscillators. Furthermore, the action of this global field on
individual oscillators is also different. 

Different models having features described above have been studied in the 
literature.
In~\cite{Hong-Strogatz-11} the contributions to the global field from all 
oscillators
were the same, but the action on the oscillators was different -- some 
oscillators were
attracted to the mean field, and some repelled from it. A generalization
of these results on the case of a general distribution of forcing strengths
is presented in~\cite{Iatsenko_etal-13}.  In paper~\cite{Paissan-Zanette-08}
the authors considered different factors for contributions to the mean field and 
for
the forcing on the oscillators, however no diversity in 
the phase shifts was studied. 
In~\cite{Montbrio-Pazo-11} only diversity of these phase shifts was considered. 

In this paper we consider a generic Kuramoto-type globally coupled model, where
all parameters of the coupling 
(factors and phases of the contributions of oscillators
to the global field, and factors and phases for the forcing of this mean field
on the individual oscillators) can be different (cf.~\cite{Iatsenko_etal-13a} 
where 
such a setup has been 
recently independently suggested). Furthermore, external noise terms
are included in the consideration. We formulate self-consistency conditions for 
the global field
and give an explicit solution of these equations in a parametric form. We 
illustrate 
the results with different cases of the coupling parameter distributions. In 
particular,
we consider a situation where the factors and phases of the coupling are 
determined by
a geometrical configuration of the oscillator distribution in space.

\section{Basic model}
\begin{figure}[h]
		\includegraphics[width=0.5\columnwidth]{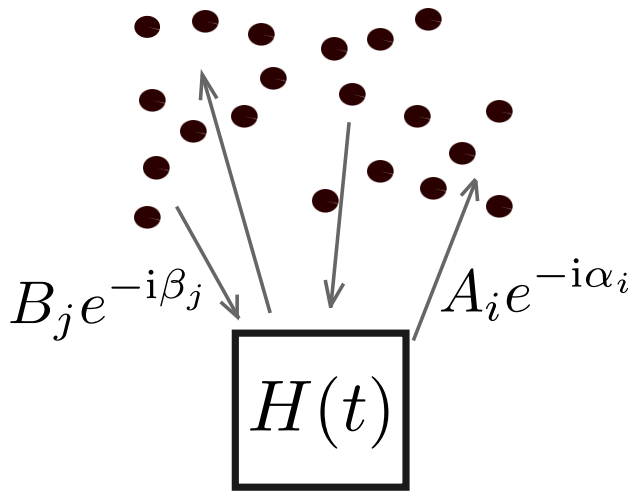}
		\caption{Configuration of the network, coupled via the mean field $H(t)$.}
		\label{fig.model}
	\end{figure}

	We consider a generic system of the Kuramoto-type phase 
  oscillators $\theta_i(t)$ having frequencies $\w_i$, with the 
  mean field coupling depicted in  Fig.~\ref{fig.model}. 
  Each oscillator $j$ contributes to 
  the mean field $H(t)$ with its own phase shift $\beta_j$ and 
  coupling constant 
  $B_j$.
	The mean field $H(t)$ acts on oscillator $i$ with a 
  specific phase shift $\alpha_i$ and a coupling strength $A_i$.
	
  It is convenient to introduce additionally the overall coupling strength $\e$
  (e.g, by normalizing one or both of the introduced quantities $A_i,B_j$; below
  for definiteness we assume $A_i,B_j>0$ because changing the sign of the  
  coupling can be absorbed to the phase shifts $\beta_j,\alpha_i$) and
  the overall phase shift $\delta$ 
  (e.g., by normalizing the shifts $\beta_j,\alpha_i$). Additionally we assume
  that the oscillators are subject to independent Gaussian white noise forces
($\langle \xi_i(t)\xi_j(t')\rangle=2\delta_{ij}\delta(t-t') $) with intensity 
$D$. 
	In this formulation the equations of motions of the oscillators read
	\be
		\label{m.1}
		\dot{\theta}_i=\omega_i + A_i {\varepsilon \over N}\sum_{j=1}^N
B_j \sin(
		\theta_j-\beta_j-\theta_i+\alpha_i-\delta)+\sqrt{D} \xi_i(t).
	\ee
	The system~(\ref{m.1}) can be rewritten in terms of the mean field
$H(t)$:
	\be
		\label{m.2}
		\begin{split}
			\dot{\theta}_i&=
      			\omega_i + A_i \, {\rm
Im}\left(H(t)e^{-\ii(\theta_i-\alpha_i)}\right)+
    		 	 \sqrt{D} \xi_i(t) , \\
			H(t)&={\varepsilon e^{-\ii \delta} \over N}\sum_{j=1}^N  B_j 
      			e^{\ii(\theta_j-\beta_j)}.
		\end{split}
	\ee
	It is convenient to reduce the number of parameters by a transformation
of phases $\vp_i=\theta_i-\alpha_i$. 
  Then the equations for $\vp_i$ are:
	\be
		\label{m.3}
		\begin{split}
			\dot{\vp}_i&=\omega_i +A_i \, {\rm
Im}\left(H(t)e^{-\ii\vp_i}\right)
      			+\sqrt{D} \xi_i(t) , \\
			H(t)&={\varepsilon  e^{-\ii \delta}\over N}\sum_{j=1}^N B_j
e^{\ii(\vp_j-\psi_j)}.
		\end{split}
	\ee
	where $\psi_j=\beta_j-\alpha_j$.
	
This model appears to be the most generic one among models of mean-field
coupled 
Kuramoto-type phase oscillators.
If all the parameters of the coupling $A_i,B_i,\beta_i,\alpha_i$ are constant,
then the model reduces to the standard Kuramoto-Sakaguchi 
one~\cite{Sakaguchi-Kuramoto-86}.
The case with different $A_i$, $\alpha_i$  and $\omega_i$ of specific form has 
been considered previously 
in refs.~\cite{Montbrio-Pazo-11,Montbrio-Pazo-11a,*Pazo-Montbrio-11}.
Also, the case with double delta distribution of $A_i$ has been studied 
in ref.~\cite{Hong-Strogatz-11}. The case $\alpha_i=\beta_i=0$ was considered 
in ref.~\cite{Paissan-Zanette-08}.
In ref.~\cite{Iatsenko_etal-13a} the 
system~(\ref{m.1}) without noise was examined. 
Below we formulate the self-consistent equation for this model and
present its explicit solution.

It should be noted that 
the complex mean field $H(t)$ is different from the ``classical'' Kuramoto order 
parameter $N^{-1}\sum_j e^{\ii\vp_j}$ and can be larger than one, depending on 
the parameters of the system. Because this mean field yields the forcing
on the oscillators, it serves as a natural order parameter for this model. 

\section{Self-consistency condition and its solution}

Here we formulate, in the spirit of the original Kuramoto approach, 
a self-consistent equation for the  mean field $H(t)$ in the thermodynamic
limit, and present its
solution.
	In the thermodynamic limit
	the quantities $\omega$, $A$, $B$ and $\psi$ have a joint distribution 
  density $g(x)=
  g(\omega, A, B, \psi)$, where $x$ is a general vector of parameters.
  While formulating in a general form, we will consider below two specific
  situations: (i) all the 
quantities  $\omega$, $A$, $B$ and $\psi$ are independent, then $g$ is a product
of four corresponding
distribution densities; 
and (ii) situation where the coupling parameters $A$, $B$, and 
$\psi$ are determined by a geometrical position of an oscillator 
and thus depend 
on this position, parametrized by a scalar parameter $x$, while 
the frequency $\w$ is distributed independently of $x$. 

Introducing the conditional probability 
  density function $\rho(\vp ,t \,|\, x)$, we can rewrite 
  the system~(\ref{m.3}) as
	\be
		\label{nf.1}
		\begin{split}
			\dot{\vp}&=\omega +A \, {\rm
Im}\left(H(t)e^{-\ii\vp}\right)+\sqrt{D} \xi(t)=\omega+A\,Q 
      \sin(\Theta -\vp)+\sqrt{D} \xi(t), \\
			H(t)&=Q e^{\ii\Theta}=\varepsilon e^{-\ii\delta}
\int g(x) B e^{-\ii
			\psi} \int_0^{2\pi} \rho(\vp ,t \,|\, x) e^{\ii\vp} d\vp
\, dx.
		\end{split}
	\ee
	It is more convenient to write equations for $\Delta\vp=\vp-\Theta$,
with the corresponding 
  conditional probability density function $\rho(\Delta\vp ,t \,|\, x)=\rho(\vp-
  \Theta ,t \,|\, x)$:
	\be
		\label{nf.2}
		\frac{d}{dt}\Delta\vp=\omega-\dot{\Theta} - A\,Q
\sin(\Delta\vp)+\sqrt{D} 
    \xi(t),
	\ee
	\be
		\label{nf.3}
		Q =\varepsilon e^{-\ii\delta} \int g(x) B e^{-\ii\psi}
\int_0^{2\pi} \rho(
		\Delta\vp ,t \,|\, x) e^{\ii\Delta\vp} d\Delta\vp \, dx.
	\ee
	The Fokker-Planck 
  equation for the conditional probability density function $\rho(\Delta\vp ,t
\,|\, x)$ follows from (\ref{nf.2}):
	\be
		\label{nf.ce}
		\pd{\rho}{t}+\pd{}{\Delta\vp}\left(\left[\omega-\dot{\Theta} -
A\,Q 
		\sin(\Delta\vp)\right]\rho \right)=D\ppd{\rho}{\Delta\vp}.
	\ee

While one cannot \textit{a priori} exclude complex regimes in
Eq.~(\ref{nf.ce}), of particular importance 
are the simplest synchronous states where  the mean field $H(t)$ rotates uniformly (this corresponds
to the classical Kuramoto solution). Therefore,
	we look for such solutions that the phase $\Theta$ of the mean field
$H(t)$ 
  rotates with a constant (yet unknown)
   frequency $\Omega$. Correspondingly, the distribution of phase differences
$\Delta\vp$ 
  is stationary in the rotating with $\Omega$ reference frame (such a solution is often called 
traveling wave):
	\be
		\label{nf.Om.1}
		\dot{\Theta}=\Omega, \ \ \dot{\rho}(\Delta\vp ,t \,|\, x)=0.
	\ee
	
	Thus, the equation for the stationary density
  $\rho(\Delta\vp ,t \,|\, x) =\rho(\Delta\vp \,|\,
x)$ 
  reads:
	\be
		\label{nf.ce.0}
		\pd{}{\Delta\vp}\left(\left[\omega-\Omega - A\,Q
\sin(\Delta\vp)\right]
		\rho \right)=D\ppd{\rho}{\Delta\vp}.
	\ee
 
 Suppose we find solution of (\ref{nf.ce.0}), which then depends on $Q$ 
 and $\W$.  Denoting  
 \be
 	\label{nf.ce.0-1}
 	F(\Omega,Q)= \int g(x) B e^{-\ii\psi} \int_0^{2\pi}
 	\rho(\Delta\vp ,t \,|\, x) e^{\ii\Delta\vp} d\Delta\vp \, dx\;,
\ee
we can then rewrite the self-consistency condition (\ref{nf.3}) as
 \be
 \label{nf.ce.0-2}
Q=\e e^{-\ii\delta} F(\Omega,Q)\;.
\ee
It is convenient to consider now $Q$, $\W$ not as unknowns but as parameters,
and to write explicit equations for the coupling strength constants $\e,\delta$
via these parameters:
\be
		\label{nf.eps-delt}
		\varepsilon=\frac{Q}{|F(\Omega,Q)|}\,, \ \ \ 
\delta=\text{arg}(F(\Omega,Q)) \,.
\ee
This solution of the self-consistency problem is quite convenient for
the numerical implementation, as it reduces to finding of solutions
of the stationary Fokker-Planck equation  (\ref{nf.ce.0}) and their integration
 (\ref{nf.ce.0-1}). Below we consider separately how this can be done 
 in the noise-free case and in presence of noise.

\section{Noise-free case}
  In the case of vanishing noise $D=0$ and (\ref{nf.ce.0}) reduces to
	\be
		\label{nf.ce.1}
		\pd{}{\Delta\vp}\left(\left[\omega-\Omega - A\,Q
\sin(\Delta\vp)\right]
		\rho \right)=0.
	\ee 
	The solution of Eq.~(\ref{nf.ce.1}) depends on the value of the
parameter $A$. 
  There are locked phases when $|A|>|\Omega-\omega|/Q$ so $\omega-\Omega -A\,
  Q \sin(\Delta\vp)=0$ 
	and rotated phases when $|A|<|\Omega-\omega|/Q$ such that
$\rho=C(A,\omega)|
  \omega-\Omega -A\,Q \sin(\Delta\vp)|^{-1}$. So the integral over parameter 
  $x$ in~(\ref{nf.ce.0-1}) splits into two integrals:
	
	\be
		\label{nf.Rh}
		\begin{split}
		F(\W,Q)&=\int_{|A|>|\Omega-\omega|/Q}g(x)Be^{-\ii\psi}\,
e^{\ii\Delta\vp(A,\omega)}  dx\,+ \\ 
&+\int_{|A|<|\Omega-\omega|/Q}g(x)Be^{-\ii\psi}\, C(A,\omega)
\int_0^{2\pi} \frac{e^{\ii\Delta\vp}\,d\Delta\vp}{|\omega-\Omega -A\,Q 
\sin(\Delta\vp)|} \,dx\,, 
\end{split}
\ee
where in the first integral
\[
\sin(\Delta\vp(A,\omega))=-\frac{\Omega-\omega}{A\,Q}\,,
\]
and in the second one
\[
C(A,\omega)=
\left(\int_0^{2\pi} \frac{d\Delta\vp}{|\omega-\Omega -A\,Q 
\sin(\Delta\vp)|}\right)^{-1}\,.
\]	
Integrations over $\Delta\vp$ in~(\ref{nf.Rh}) can be performed explicitely:
	\be
		\label{nf.int}
		\begin{split}
C(A,\omega)=&\left(\int_0^{2\pi} 
\frac{d\Delta\vp}{|\omega-\Omega -A\,Q \sin(\Delta\vp)|}\right)^{-1}=
\frac{\sqrt{(\Omega-\omega)^2-A^2Q^2}}{2\pi}\, , \\
&\int_0^{2\pi} \frac{e^{\ii\Delta\vp}\,d\Delta\vp}{|\omega-\Omega -A\,Q 
\sin(\Delta\vp)|}=\frac{2\pi\ii}{AQ}
\left(\frac{\Omega-\omega}{|\Omega-\omega|}-
\frac{\Omega-\omega}{\sqrt{(\Omega-\omega)^2-A^2Q^2}}\right) \, .
		\end{split}
	\ee
	After substitution~(\ref{nf.int}) into~(\ref{nf.Rh}), we 
  obtain the final general expression for the main function  $F(\W,Q)$:
	\be
		\label{nf.Rh.1}
		\begin{split}
F(\W,Q)&=\int_{|A|>|\Omega-\omega|/Q}g(x)Be^{-\ii\psi}\, 
      \sqrt{1-\frac{(\Omega-\omega)^2 }{A^2 Q^2 } \,}\, dx\, - \\
			&-\ii \int
g(x)Be^{-\ii\psi}\,\frac{\Omega-\omega}{A\,Q}\, dx \, + \\
			&+\ii \int_{|A|<|\Omega-\omega|/Q}g(x)Be^{-\ii\psi}\,
      \frac{\Omega-\omega}{|\Omega-\omega|}
      \sqrt{\frac{(\Omega-\omega)^2 }{A^2 Q^2 }-1 \,}  \ dx\, .
		\end{split}
	\ee

\subsection{Independent parameters}
	
The integrals in~(\ref{nf.Rh.1}) simplify in the case of 
independent distributions of the parameters $\omega$, $A$ and $B$, $\psi$. That 
means that 
$g(x)=g_1(\omega,A)\,g_2(B,\psi)$. In this case it is convenient to consider
$\e$ and $\delta$ as scaling parameters of the
distribution $\tilde{g}_2(\tilde{B},\tilde{\psi})$, such that
\be
	\label{nf.Bpsi.1}
	\e e^{-\ii\delta}=\int\int
\tilde{g}_2(\tilde{B},\tilde{\psi})\tilde{B}e^{-\ii\tilde{\psi}}
d\tilde{B}d\tilde{\psi},
\ee
so the parameters $B=\tilde{B}/\e$ and $\psi=\tilde{\psi}-\delta$  have such a
distribution $g_2(B,\psi)=\e \tilde{g}_2(\tilde{B},\tilde{\psi})$ that satisfies
\be
	\label{nf.Bpsi.2}
	\int\int g_2(B,\psi)Be^{-\ii\psi} dBd\psi=1.
\ee
From Eq.~(\ref{nf.Bpsi.2}) it follows that Eq.~(\ref{nf.Rh.1}) reduces,
because the integration over $B$ and $\psi$ yields $1$, to the following expression:
	\be
		\label{nf.Rh.2}
		\begin{split}
			F(\W,Q)&=
\int\int_{|A|>|\Omega-\omega|/Q}g_1(\omega,A)\,
\sqrt{1-\frac{(\Omega-\omega)^2 }{A^2 Q^2 } \,}\, dAd\omega\, - \\
			&-\ii \int\int
g_1(\omega,A)\frac{\Omega-\omega}{A\,Q}\, dAd\omega \, +\\
&+\ii
\int\int_{|A|<|\Omega-\omega|/Q}g_1(\omega,A)\,\frac{\Omega-\omega}{
|\Omega-\omega|}\sqrt{\frac{(\Omega-\omega)^2 }{A^2 Q^2 }-1 \,}  \ dAd\omega\,
.
		\end{split}
	\ee
	Then  the parameters $\e$ and $\delta$ can be found from 
Eqs.~(\ref{nf.eps-delt}) depending on $\Omega$ and $Q$.   Noteworthy, 
 all the complexity of distributions of parameters $B$ and $\psi$ is accumulated 
 in values of $\e$ and $\delta$, while distributions of $\w,A$ still 
 contribute to the integrals.

\begin{figure}[h]
		(a)\includegraphics[width=0.45\columnwidth]{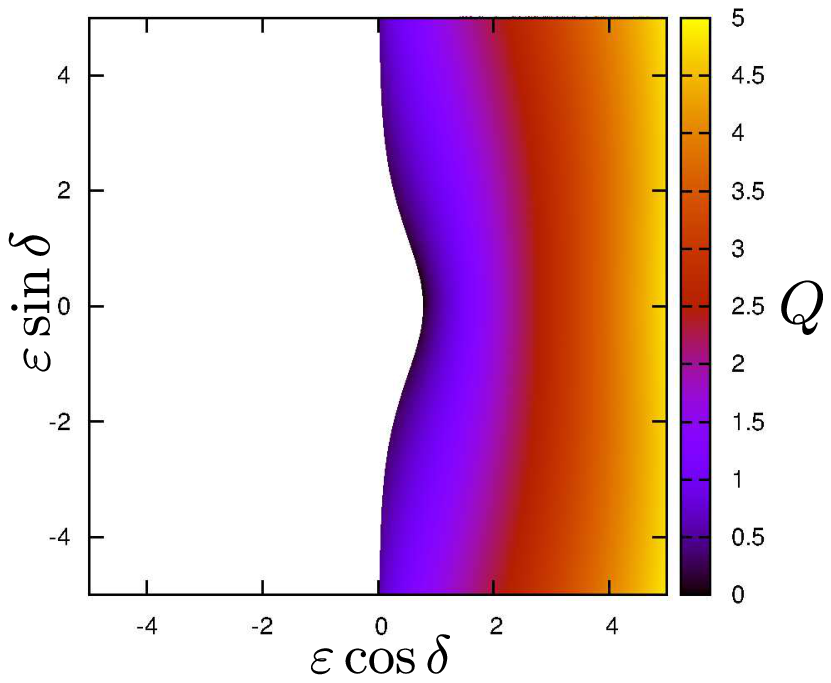}\hfill
(b)\includegraphics[width=0.45\columnwidth]{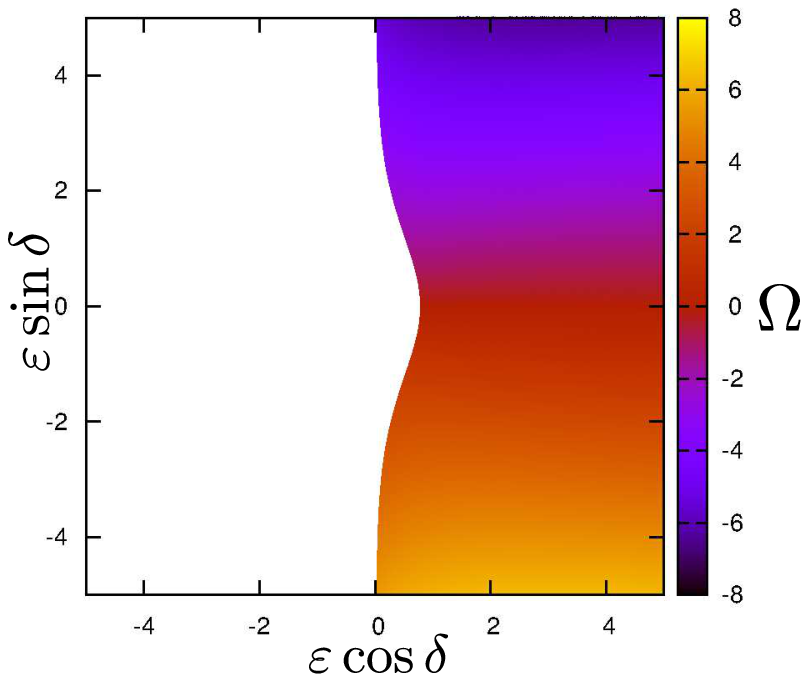}
		\caption{(color online) Dependencies of the amplitude $Q$
of the mean field (a) and of its frequency $\Omega$  on the
parameters $\e$ and $\delta$, for $\theta=1$. White area corresponds to
asynchronous state with vanishing mean field.}
		\label{fig.w_R_thpsi}
	\end{figure}

Below we give an example of application of our theory.
In Fig.~\ref{fig.w_R_thpsi} we present results of the calculation 
of the order parameter $Q$ and the frequency of the global field $\Omega$
as  function  $\e,\delta$ , for $g_1(\omega,A)=g(A)g(\omega)$ and
$g(A)=\frac{A}{\theta^2}e^{-A/\theta}$, $g(\omega)=\frac{1}{\sqrt{2\pi}}
e^{-\omega^2 /2}$.

	Furthermore, Eq.~(\ref{nf.Rh.2}) simplifies even more when the individual
frequencies of the oscillators are identical, i.e.  when
$g(\w)=\delta(\w-\w_0)$. Then the integration over $d\omega$ can be performed
first:
	\be
		\label{nf.Rh.3}
		\begin{split}
			F(\W,Q)&= \int_{|A|>|\Omega-\omega_0|/Q}g(A)\,
\sqrt{1-\frac{(\Omega-\omega_0)^2 }{A^2 Q^2 } \,}\, dA\, - \\
			&-\ii \int g(A)\frac{\Omega-\omega_0}{A\,Q}\, dA \,
+\ii
\int_{|A|<|\Omega-\omega_0|/Q}g(A)\,\frac{\Omega-\omega_0}{|\Omega-\omega_0|}
\sqrt{\frac{(\Omega-\omega_0)^2 }{A^2 Q^2 }-1 \,}  \ dA\, .
		\end{split}
	\ee
	It is convenient to treat the function $F(\W,Q)$ in
Eq.~(\ref{nf.Rh.3}) as a function of a new variable
$Y=\frac{\Omega-\omega_0}{Q}$, which is a combination of variables $\W$ and $Q$. Then
Eq.~(\ref{nf.Rh.3}) for $F(\W,Q)$ transforms to the following equation for
$F(Y)$
	\be
		\label{nf.Rh.4}
		\begin{split}
			F(Y)&= \int_{|A|>|Y|}g(A)\, \sqrt{1-\frac{Y^2 }{A^2}
\,}\, dA\, - \\
			&-\ii \int g(A)\frac{Y}{A}\, dA \, +\ii
\int_{|A|<|Y|}g(A)\,\frac{Y}{|Y|}\sqrt{\frac{Y^2}{A^2}-1 \,}  \ dA\, ,
		\end{split}
	\ee
where we took into account that $Q\geq0$.	

	Despite the fact that Eqs~(\ref{nf.eps-delt}) are still valid for
finding $\e$ and $\delta$, it is more convenient to use $Y$ and $\e$ as a parameters in
Eq.~(\ref{nf.ce.0-2}) instead of $Q$ and $\W$. Then the final expressions
  for finding $Q$, $\W$ and $\delta$ take the following form:
	\be
		\label{nf.eps-delt.1}
		Q=\e\, |F(Y)|\,, \ \ \  \W=\w_0+ \e Y\, |F(Y)|\,, \ \ \
\delta=\text{arg}(F(Y)) \,.
	\ee
	
\begin{figure}[h]
		(a)\includegraphics[width=0.45\columnwidth]{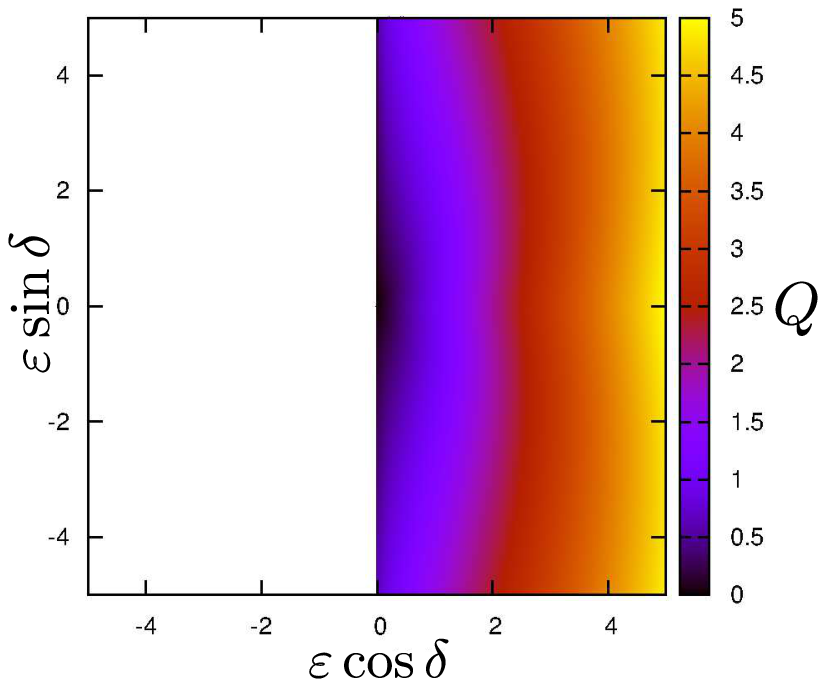}\hfill
(b)\includegraphics[width=0.45\columnwidth]{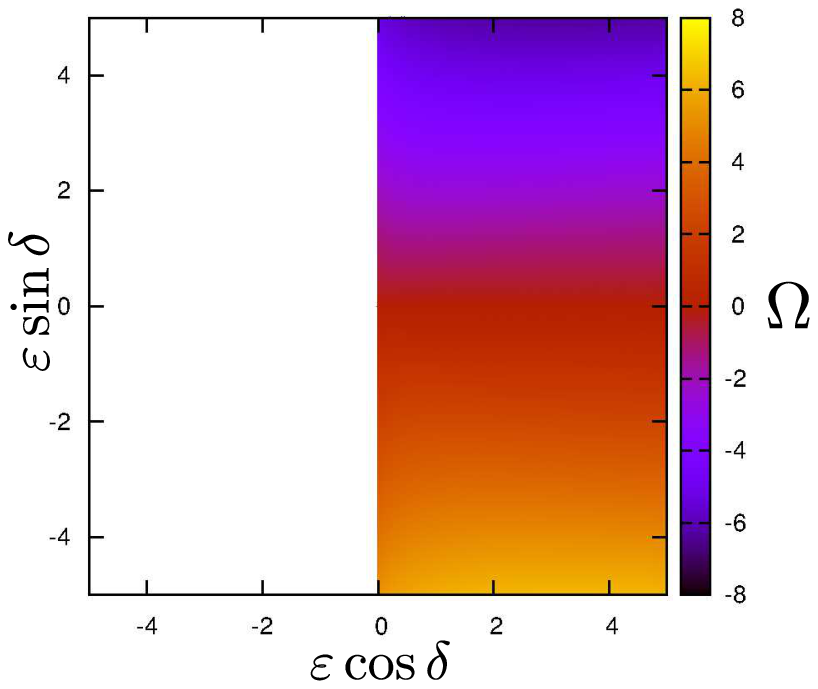}
		\caption{(Color online) Dependencies of the amplitude $Q$
of the mean field (a) and of its frequency $\Omega$  on the
parameters $\e$ and $\delta$, for $\theta=1$ and $\w_0=0$.
 White area corresponds to
asynchronous state with vanishing mean field.}
		\label{fig.deltw_R_thpsi}
	\end{figure}

	The results of the calculation of $Q(\e,\delta)$ and
$\Omega(\e,\delta)$ for the identical natural frequencies are shown in
Fig.~\ref{fig.deltw_R_thpsi}, where we chose 
$g_1(\omega,A)=\frac{A}{\theta^2}e^{-A/\theta}\delta(\w-\w_0)$.

  Summarizing this section, we have presented general expressions for the order 
parameter, frequency of the mean field  and the coupling parameters in a 
parametric form. These expressions are exemplified for specific distributions of 
the strengths and phase shifts in the couplings in Figs.~\ref{fig.w_R_thpsi},
\ref{fig.deltw_R_thpsi}. In the case of a distribution of natural frequencies 
(Fig.~\ref{fig.w_R_thpsi}) there is a threshold in the coupling for the onset of 
collective dynamics. For the oscillators with equal frequencies  
(Fig.~\ref{fig.deltw_R_thpsi}) there is no threshold.

\section{Self-consistent solution in the presence of noise}

Here we have to find the stationary solution of the Fokker-Planck equation
(\ref{nf.ce.0}). It
 can be solved in the Fourier modes representation:
	\be
		\label{n.Fm}
		\rho(\Delta\vp \,|\, x)=\frac{1}{2\pi}\sum_n C_n(x) e^{\ii n
\Delta\vp}\;,
\qquad C_n(x)=\int_0^{2\pi}\rho e^{-\ii n\Delta\vp}d\Delta\vp\;,\qquad C_0(x)=
1\; .
	\ee
Substituting (\ref{n.Fm}) in Eq.~(\ref{nf.ce.0}),   we obtain
	\be
		\label{n.eq.Fm}
		\begin{split}
			&\int_0^{2\pi}d\Delta\vp
			\left[-\frac{\partial}{\partial
\Delta\vp}([\omega-\Omega-AQ
\sin(\Delta\vp)]\rho)+D\frac{\partial^2\rho}{\partial\Delta\vp^2}\right]e^{-\ii
k\Delta\vp}=\\
			&=-k^2DC_k+\ii k(\Omega -\omega)C_k+\ii
kAQ\frac{C_{k-1}-C_{k+1}}{2\ii}=0.
		\end{split}
	\ee
	As a consequence, we get a tridiagonal system of 
  algebraic equations
	\be
		\label{n.eq.1}
		[2kD-\ii2(\Omega -\omega)]C_k+AQ(C_{k+1}-C_{k-1})=0.
	\ee
	The infinite system~(\ref{n.eq.1}) can be solved by 
  cutting it at some large $N$, as follows (see ref.~\cite{Risken-89}):
	\be
		\label{n.eq.2}
		C_k=\alpha_k C_{k-1}, \quad a_k=2kD-\ii2(\Omega -\omega)\;,\quad
			\alpha_N=\frac{AQ}{a_N}\;,\quad 
			\alpha_k=\frac{AQ}{a_k+AQ\alpha_{k+1}}.
	\ee
	As a result, $C_1$ can be found recursively as a continued fraction:
	\be
		\label{n.eq.4}
		C_1=\alpha_1=\frac{AQ}{a_1+AQ\alpha_{2}}=\ldots\;.
	\ee
	From Eq.~(\ref{n.eq.4}) it is obvious that in general $C_1$ is a
function of $\Omega$, $Q$, $\omega$ and $A$:
	\be
		\label{n.C1}
		C_1=C_1(\Omega, Q, \omega, A).
	\ee
	
	The integral over $\Delta\vp$ in~(\ref{nf.ce.0-1}) can be calculated
using the Fourier-representation~(\ref{n.Fm}), yielding
	\be
		\label{n.int.Rh}
		\int_0^{2\pi} \rho(\Delta\vp \,|\, x) e^{\ii\Delta\vp}
d\Delta\vp = C_1^*(\Omega, Q, \omega, A) .
	\ee
	Thus the expression for $F$ in the case of noisy oscillators reads
	\be
		\label{n.F.0}
		F(\Omega,Q)=\int g(x) B e^{-\ii\psi} C_1^*(\Omega, Q,
\omega, A) dx.
	\ee
	
  \subsection{Independent parameters}
  
	From the expression~(\ref{n.C1}) it follows that the integral
in~(\ref{n.F.0}) 
  simplifies in the same case of independent distribution of the parameters 
  $g(x)=g_1(\omega,A)\,g_2(B,\psi)$, similar to the noise-free case
  described in the previous section. Here we use the 
  same notations as before, including condition~(\ref{nf.Bpsi.2}).
  
\begin{figure}
(a)		\includegraphics[width=0.45\columnwidth]{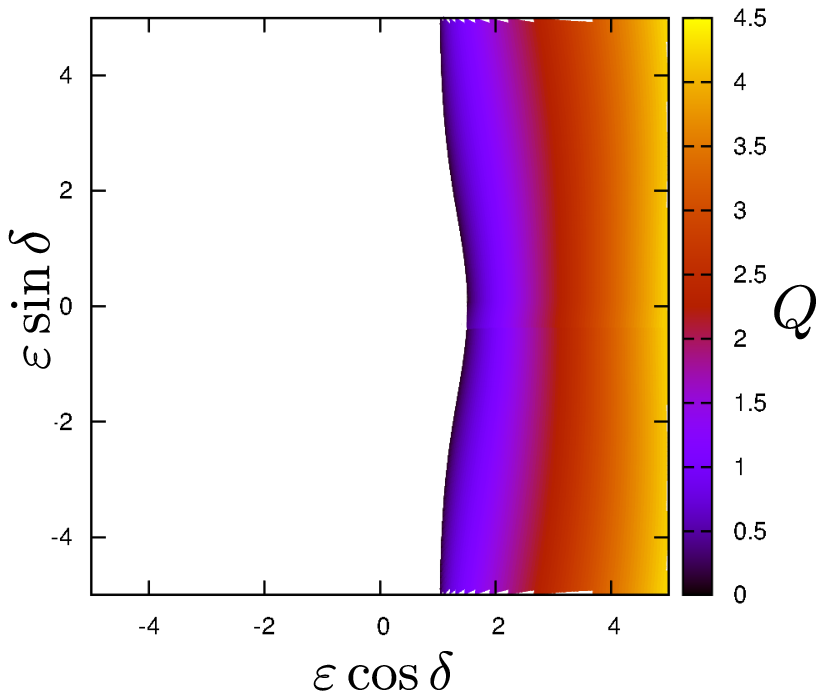}\hfill
(b)\includegraphics[width=0.45\columnwidth]{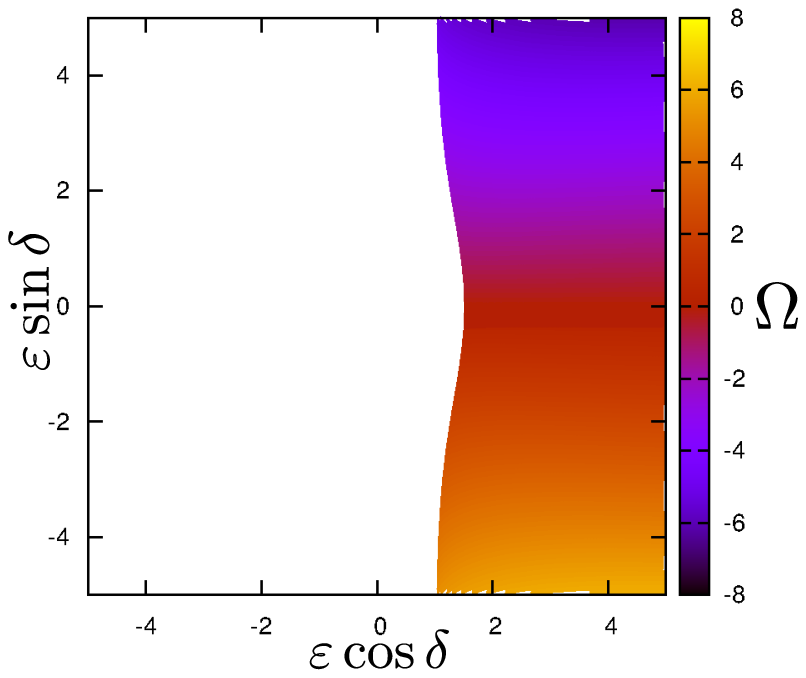}
		\caption{(Color online) The same as Fig.~\ref{fig.w_R_thpsi}, but with noise $D=1$}
		\label{fig.w_R_thpsi_noise}
	\end{figure}

  The parameters $\e$ and $\delta$  can be 
  found from  Eqs.~(\ref{nf.eps-delt}), where $F(\Omega, Q)$ is determined
from
	\be
		\label{n.F}
		F(\Omega,Q)=\int g_1(\omega,A) C_1^*(\Omega, Q, \omega, A)
dAd\omega .
	\ee
  In this way we obtain $Q(\e,\delta)$ and
$\Omega(\e,\delta)$ (Fig.~\ref{fig.w_R_thpsi_noise}). For calculations we used the same 
   distribution $g_1(\omega,A)$ as in the noise-free case.

\begin{figure}[h]
		(a) \includegraphics[width=0.45\columnwidth]{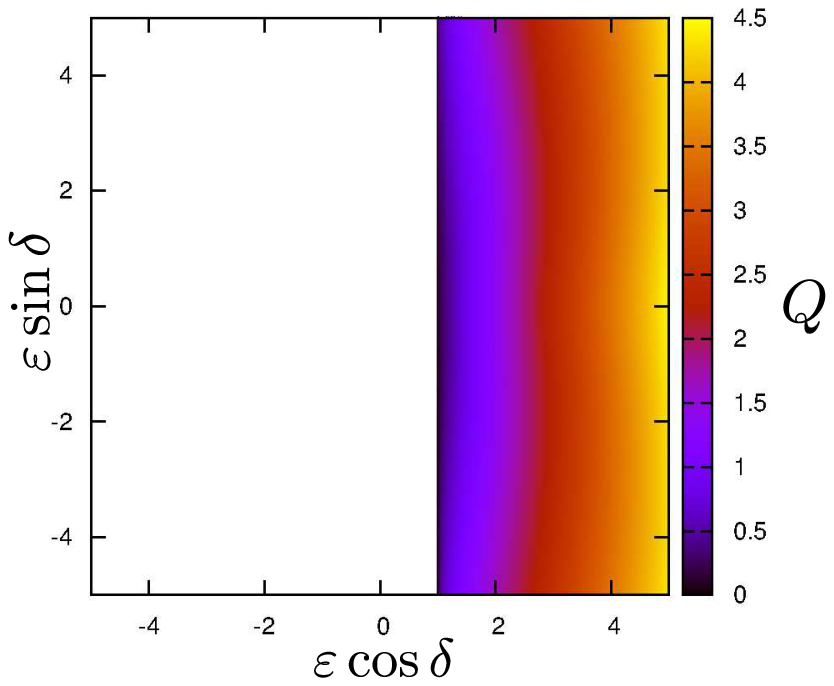}\hfill
(b) \includegraphics[width=0.45\columnwidth]{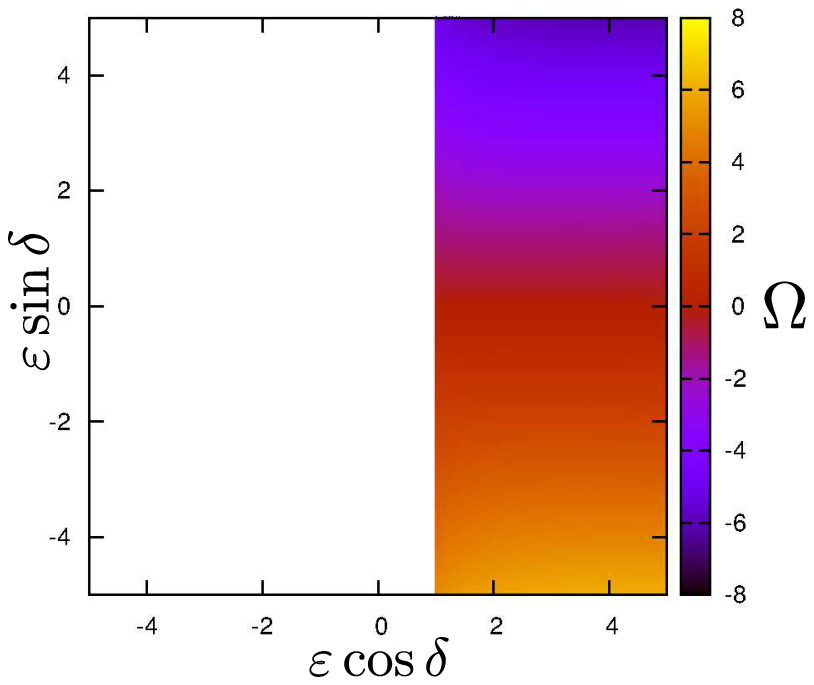}
		\caption{(Color online)The same as Fig.~\ref{fig.deltw_R_thpsi}, 
but with noise $D=1$.
}
		\label{fig.deltw_R_thpsi_noise}
	\end{figure}

Contrary to the noise-free case, when oscillator's individual 
frequencies are identical (delta-function distribution), 
no further simplification of $F(\W,Q)$ appears possible. 
In 
Fig.~\ref{fig.deltw_R_thpsi_noise} we report the results for the same parameters 
as in 
Fig.~\ref{fig.deltw_R_thpsi}, but with noise $D=1$.
	
In the considered model the main effect caused by noise 
is the shift of the synchronization threshold to larger values of 
the coupling strength $\e$. The noise acts very much similar to the distribution
of natural frequencies;  if the oscillator's individual frequencies are identical,
noise leads to a non-zero threshold 
in the coupling. 
	
\section{Example of a geometric organization of oscillators}

	In this section we present a particular example of  application
of general expressions above to the case where distributions of parameters
are determined by configuration of oscillators.  
We consider
spatially spreaded phase oscillators with
a common receiver that collects signals from all oscillators, and with an 
emitter
that receives the summarized signal from the receiver and sends 
the coupling signal to the oscillators; below we assume that these
signals
  propagate with velocity $c$. We assume that the oscillators have the same natural
frequencies $\omega_0=1$ (cases where the frequencies are distributed (dependent or independent of geometric positions of oscillators) can be straightforwardly treated within the same framework).

\begin{figure}
		\includegraphics[width=0.6\columnwidth]{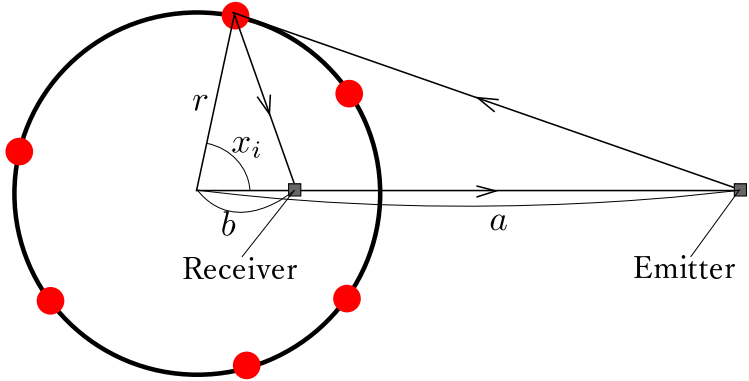}
		\caption{(Color online) The scheme of the system.}
		\label{fig.scheme}
	\end{figure}

	We assume that oscillators are distributed uniformly on a circle of
radius $r$. Each oscillator is thus labeled by the angle $x_i$
(Fig.~\ref{fig.scheme}).
	The receiver, the emitter, and the center of the circle are supposed to 
  lie on one line.

	Also, we assume that the phase shifts $\beta_j$ and $\alpha_i$ are
proportional to the distances between the oscillator, the receiver and the 
emitter, so that
the system can be described by Eq.~(\ref{m.1}), where:	
	\be
		\label{ex.beta.1}
		\beta_j={\omega_0 \over c} \sqrt{r^2+b^2-2rb\cos x_j}\,,\quad
		\alpha_i={\omega_0 \over c} \sqrt{r^2+a^2-2ra\cos x_i}\,.
	\ee
	Coupling strengths $B_j$ and $A_i$ are inversely proportional to the
square distances between each oscillator, receiver and emitter:
	\be
		\label{ex.B.1}
		B_j=\frac{1}{r^2+b^2-2rb\cos x_j},\quad
		A_i=\frac{1}{r^2+a^2-2ra\cos x_i}\;,
	\ee
where $a$ and $b$ is the distances from the center of the circle to the emitter 
and the receiver respectively (Fig.~\ref{fig.scheme}).
	The parameters $\e$ and $\delta$ can be interpreted as a coupling
coefficient and a phase shift for the signal transfer from the receiver to the
emitter.

The theory developed above yields stable solutions for 
any given parameters $a$ and $b$.
	Since all the oscillators have the same natural frequencies, the variable 
  transformation
	$Y=(\Omega-\w_0)/Q$ described in section III should be performed. Thus, it is 
  suitable to use Eqs.~(\ref{nf.eps-delt.1}) in order to find $Q$, $\Omega$ and 
  $\delta$ as a functions of $\e$ and $Y$.

\begin{figure}
		\includegraphics[width=0.45\columnwidth]{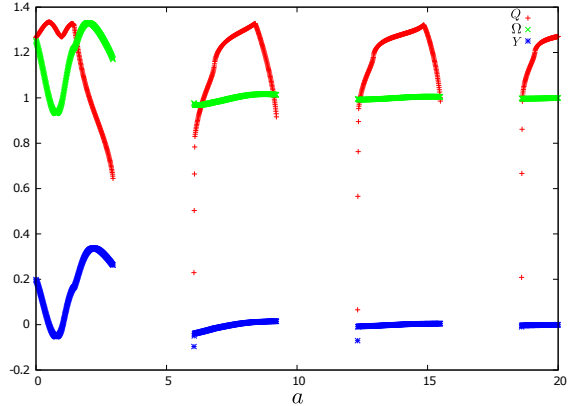}
		\caption{(Color online) The dependance of $Q(a)$, $\Omega(a)$ and $Y(a)$ on the distance from the center of the circle to the emitter $a$, $b=0.5$, $r=1$, $\e=1$ and $\delta=0$. Periodicity in $a$ corresponds to the zones of attractive and repulsive coupling due to delay-induced phase shift.}
		\label{fig.circle_Q_a}
	\end{figure}

In the numerical example presented in Fig.~\ref{fig.circle_Q_a}, 
we  fixed $b=r/2$ and varied $a$,  
finding the order parameter $Q(a)$ and  the frequency of the collective 
oscillations
$\Omega(a)$ for $\e=1$ and $\delta=0$.  One can see a sequence of 
synchronization regions separated by asynchronous intervals; this
is typical for systems with time delay in the coupling -- in our case this
delay is due to separation of the emitter from the community of oscillators, and 
the finite speed of signal propagation assumed.  The dependencies shown are not 
smooth, because as parameter $a$ varies, some oscillators enter/leave the 
synchronization domain.

\section{Conclusion}
We have developed a theory of synchronization for phase oscillators 
on networks with a special structure of coupling through a global mean field.
A unified description of the frequency and the amplitude of the mean field 
in a parametric form is valid both for noise-free and noisy oscillators. In the 
latter case numerical evaluation of a 
continued fraction is needed, otherwise the
solution reduces to calculation of integrals over parameter distributions. As 
one of the examples we considered a situation, where contributions to the mean 
field and its action on oscillators are prescribed by a geometric configuration
of the oscillators; phase shifts and the contribution factors result from the
propagation of the signals as waves having certain velocity. The general 
formulation we 
developed can be used for any such configuration. It appears that the method
above may be useful also in more general network 
setups, where there is no global
mean field, but such a field can be introduced as approximation 
(cf.~\cite{Skardal-Restrepo-12,Burioni-etal-13}).

\begin{acknowledgments}
V. V. thanks the IRTG 1740/TRP 2011/50151-0, funded by the DFG /FAPESP.
\end{acknowledgments}

%

\end{document}